\documentclass[a4paper,11pt]{article}
\pdfoutput=1 % if your are submitting a pdflatex (i.e. if you have
\usepackage{jinstpub} % for details on the use of the package, please
\usepackage{lineno}
\usepackage{amsmath}
\usepackage{xcolor}

\title{\boldmath Characterisation of Gamma-irradiated MCz-Silicon Detectors with a High-$K$ Negative Oxide as Field Insulator}

\author[a,b,1]{S. Bharthuar,\note{Corresponding author.}}
\author[a,c]{M. Bezak,}
\author[a,b]{E. Brücken,}
\author[f]{A  Gädda,}
\author[a,c]{M. Golovleva,}
\author[a,c,e]{A. Karadzhinova-Ferrer,}
\author[c]{A. Karjalainen,}
\author[a,c]{N. Kramarenko,}
\author[a,b]{S. Kirschenmann,}
\author[a,c]{P. Luukka,}
\author[d]{J. Ott,}
\author[a,b]{E. Tuominen,}
\author[a,c]{M. Väänänen.}
\affiliation[a]{Helsinki Institute of Physics, Gustaf H\"{a}llstr\"{o}min katu 2,  FI-00014,  Finland}
\affiliation[b]{Department of Physics, University of Helsinki, Gustaf H\"{a}llstr\"{o}min katu 2,  FI-00014, Finland}
\affiliation[c]{Lappeenranta-Lahti University of Technology, Yliopistonkatu 34, Lappeenranta, FI-53850, Finland}
\affiliation[d]{SCIPP, University of California Santa Cruz, CA 95064, USA}
\affiliation[e]{Ludong University, Yantai, China}
\affiliation[f]{Okmetic Oy., Piitie 2, Vantaa, FI-01510, Finland}
\emailAdd{shudhashil.bharthuar@cern.ch}

\abstract{ The high-luminosity operation of the Tracker in the Compact Muon Solenid (CMS) detector at the Large Hadron Collider (LHC) experiment calls for the development of silicon-based sensors. This involves implementation of AC-coupling to micro-scale pixel sensor areas to provide enhanced isolation of radiation-induced leakage currents. The motivation of this study is the development of AC-pixel sensors with negative oxides (such as aluminium oxide - Al$_2$O$_3$ and hafnium oxide - HfO$_2$) as field insulators that possess good dielectric strength and provide radiation hardness. Thin films of Al$_2$O$_3$ and HfO$_2$ grown by atomic layer deposition (ALD) method were used as dielectrics for capacitive coupling. A comparison study based on dielectric material used in MOS  capacitors indicate HfO$_2$ as a better candidate since it provides higher sensitivity (where, the term sensitivity is defined as the ratio of the change in flat-band voltage to dose) to negative charge accumulation with gamma irradiation.

Further, space charge sign inversion was observed for sensors processed on high resistivity p-type Magnetic Czochralski silicon (MCz-Si) substrates that were irradiated with gamma rays up to a dose of 1~MGy. The inter-pixel resistance values of heavily gamma irradiated AC-coupled pixel sensors suggest that high-$K$ negative oxides as field insulators provide a good electrical isolation between the pixels.
}


\arxivnumber{1234.56789} % only if you have one

\proceeding{23$^{\text{rd}}$ International Workshop on Radiation Imaging Detectors\\
  26–30 June 2022 \\
  Riva del Garda, Italy}

\begin{document}
\maketitle 
\flushbottom

\section{Introduction}
\label{sec:intro}

%The Large Hadron Collider (LHC) experiment aims at studying elementary particles that make up matter, as explained in the Standard Model of particle physics. The major experiments in the LHC comprise of calorimeters to measure energy of the particles, particle-identification detectors to pin down the nature of the particle and tracking devices to calculate the curvature in the trajectory of a charged particle in order to determine its momentum. 
Silicon-based particle detectors are implemented in the Large Hadron Collider (LHC) experiment for vertex and track reconstruction of charged particles. However, with the upgrade of the LHC to High-luminosity LHC (HL-LHC), pixel detectors, in particular those located close to the interaction point, will be subjected to a radiation-hard environment. This leads to a degradation of their electrical properties due to radiation-induced damage~\cite{Moll}. For instance, in the HL-LHC operation of the CMS Tracker detector, the innermost silicon layers will be exposed to radiation levels increasing up to a magnitude of > 2$\times$10$^{16}$ $n_{eq}$cm$^{-2}$, and 12 MGy~\cite{Orfanelli}.

\begin{figure}[htbp]
\centering % \begin{center}/\end{center} takes some additional vertical space
\includegraphics[width=1\textwidth]{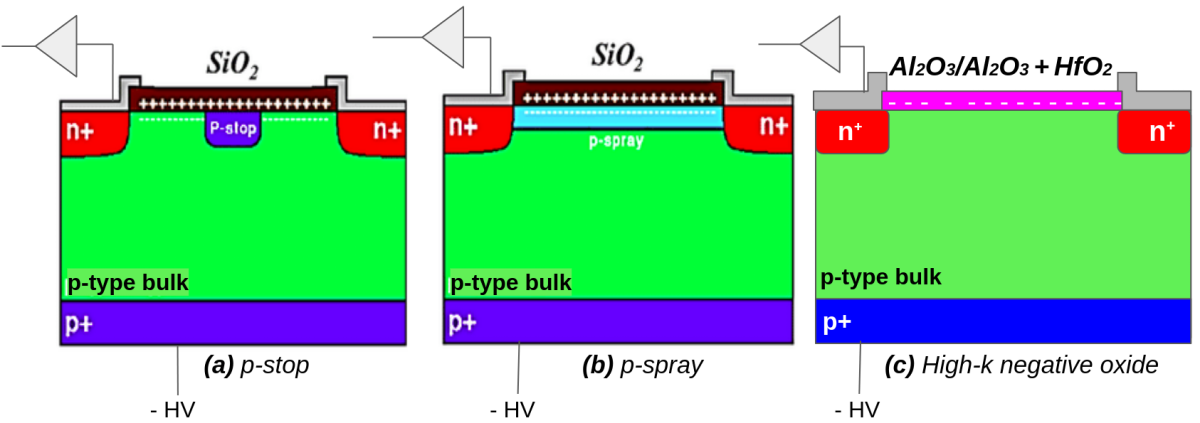}
% "\includegraphics" from the "graphicx" permits to crop (trim+clip)
% and rotate (angle) and image (and much more)
\caption{\label{oxide}Schematic representation showing electrical isolation between two adjacent pixels in a DC-coupled pixel detector where \textbf{(a)} p-stop, \textbf{(b)} p-spray or \textbf{(c)} a high-$K$ negative oxide material like Al$_2$O$_3$ or Al$_2$O$_3$ + HfO$_2$ is used as an inter-pixel isolation strategy.  }
\end{figure}

These high doses of radiation levels deteriorate the performance of the silicon detectors. This includes an decrease in their signal-to-noise ratio due to an increase in the leakage current as well as a reduction in the charge collection efficiency due to radiation induced defects. Thus, in order to alleviate these challenges, unlike the traditionally used DC-coupled (conductively-coupled) detectors that are susceptible to an increased level in leakage current values, AC-coupled (capacitively-coupled) pixel detectors are anticipated to be operational as particle tracking devices in future collider experiments. 
%Further, the implementation of AC-coupling to micro-scale pixel areas enables in providing an enhanced inter-pixel isolation to radiation-induced leakage currents. 
Conventionally, as shown in Figures~\ref{oxide}a and \ref{oxide}b, the pixels are electrically isolated from each other by the use of p-stop or p-spray in order to avoid the creation of a short-circuiting channel between the pixelated n$^+$-implants that could be generated due to the accumulation of electrons under the silicon-dioxide (positive oxide) dielectric layer. However, introduction of p-stop and p-spray requires additional implantation processing steps, which subject the silicon wafers to high temperatures and increases the mask levels that reduce the cost-effectiveness of the finally processed detectors. Alternatively as a solution, shown in Figure~\ref{oxide}c, the segmented implants can be electrically isolated from each other by utilising Atomic Layer Deposition (ALD)-grown thin films of negative oxides with a high $K$-value (Al$_2$O$_3$ and HfO$_2$) as insulating layers that possess high negative oxide charge concentration values of the order of magnitude of 10$^{11}$--10$^{13}$~cm$^{-2}$. This corroborates higher oxide capacitance values for an improved capacitive coupling~\cite{Harkonen}. ALD ensures in providing high uniformity thin films of Al$_2$O$_3$ and HfO$_2$ with good accuracy, deposited at comparatively low temperatures (typically $\sim$300$^{\circ}$C). This prevents the wafers from being subjected to additional mask levels and high temperature implantation processes.

\section{Samples Measured and Methods}
\label{sec:samplesmeasured}
All the samples were fabricated on p-type (boron-doped) Magnetic Czochralski (MCz) 6” silicon wafers from Okmetic Oy. The silicon wafers possess a resistivity and thickness of $\sim$4-8~k$\Omega$-cm and 320~\textmu m, respectively. Fabrication of the devices measured in this study have been provided in articles~\citep{ott1, gadda, ott2}.

 The following study focuses on investigating the robustness and the impact of mostly gamma irradiation, up to a dose of 1~MGy, on ALD-grown Al$_2$O$_3$ or HfO$_2$ + Al$_2$O$_3$ implemented as a field insulator in Metal-Oxide-Semiconductor (MOS) capacitors (with a gate of 1.5~mm in diameter), as well as in capacitive coupling for n$^+$/p$^-$/p$^+$ prototypes of PIN diode (where the active area of the pad is 7.2~$\times$~7.2~mm$^2$) and AC-coupled pixel detectors (similar to the PSI46dig design consisting of 52~$\times$~80 pixel matrix, along with a pitch size of 150~$\times$~100~\textmu m)~\cite{2013JInst...8C2047G}. Previous studies in~\cite{bharthuar} show that an additional layer of HfO$_2$ along with Al$_2$O$_3$ incorporated in MOS devices provides a higher capacitive coupling along with a better insulation, reduced susceptibility to an early breakdown, higher sensitivity to proton irradiation specifically at high fluences, and an improved radiation hardness. Additionally, electrical characterisation studies were performed on the non-irradiated and gamma-irradiated sensors to study the electric field within the bulk of PIN diode prototypes and inter-pixel resistance in AC-coupled pixel sensors. 

%Additionally, defect characterisation of the sensors with high-$K$ negative oxide (specifically Al$_2$O$_3$) were carried out using Deep-Level Transient Spectroscopy (DLTS) measurements to identify the defects introduced in the detector either during the silicon crystal growth or the processing steps.
 
\section{Characterisation of Gamma Irradiated MOS Capacitors with high-$K$ negative oxides}

\label{sec:Moscap}
Capacitance-versus-Voltage measurements (C-V) of MOS capacitors can be used to determine the effective fixed oxide and mobile charges of the insulating layer by extracting the flat-band voltage ($V_\text{fb}$). $V_\text{fb}$ is defined as the gate voltage value at which the the energy band of the silicon substrate is flat at the oxide-substrate interface. In ideal conditions, $V_\text{fb}$ is of a magnitude of -0.54~V for MOS devices with aluminum as the metal gate and a silicon substrate with a doping concentration of 8~$\times$~10$^{11}$~cm$^{-3}$. The shift in the flat-band voltage ($\Delta V_\text{fb}$) value from its ideal condition gives an estimation of effective fixed oxide charge density value ($N_\text{f}$) in the dielectric layer that are incorporated during the thin films growth or deposition processes.
\begin{figure}[htbp]
\centering
\includegraphics[width=1.03\textwidth]{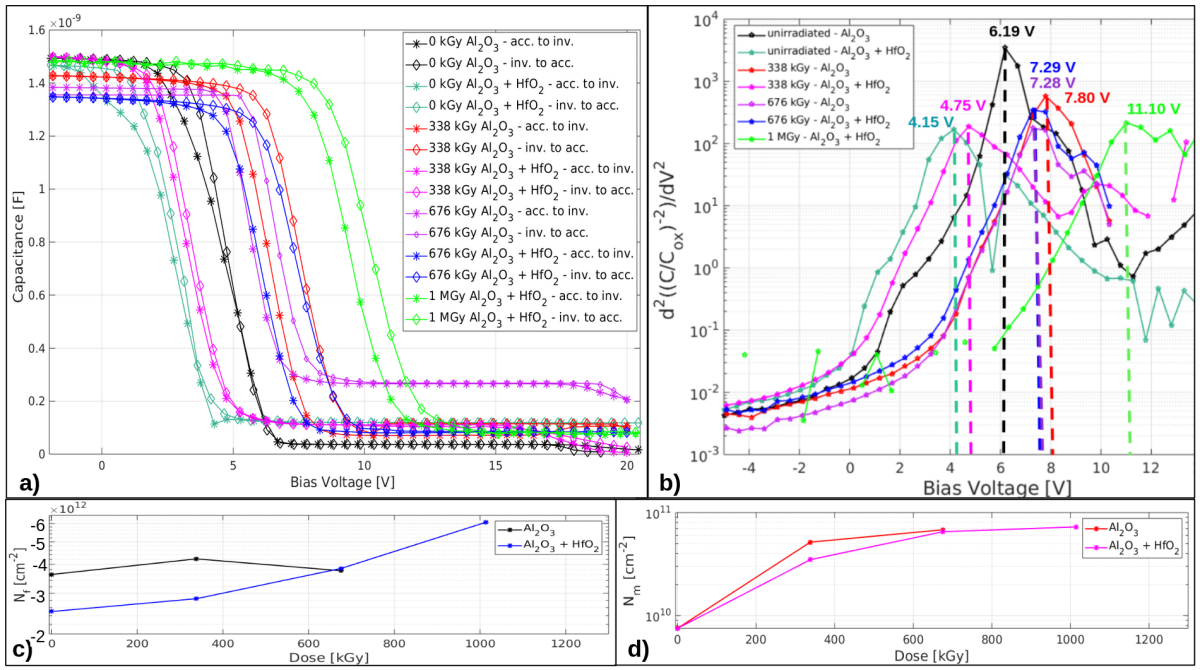}
\caption{\label{fig:2}\textbf{a)}~C-V curves for non-irradiated and gamma-irradiated MOS capacitors with Al$_2$O$_3$ and Al$_2$O$_3$+HfO$_2$ as the dielectric layer, measured by scanning the bias voltage from accumulation to inversion mode of operation and vice-versa at a constant frequency of 1~kHz. \textbf{b)}~Estimation of $V_\text{fb}$ from C-V curves, where the bias voltage is scanned from accumulation to inversion region, as the peak bias voltage value, extracted by double-differentiating $C$/$C_\text{ox}$ curve with respect to bias voltage. Variation in concentration of effective \textbf{c)}~fixed oxide charges ($N_\text{f}$) and \textbf{d)}~mobile oxide charges ($N_\text{m}$) with gamma dose.}
\end{figure}
Figure~\ref{fig:2}a shows the C-V curves for MOS capacitors with either Al$_2$O$_3$ or Al$_2$O$_3$+HfO$_2$ as the oxide layer irradiated with gamma-rays at doses of 338~kGy, 676~kGy and 1~MGy. The capacitance values were recorded with an LCR meter at a frequency of 1~kHz and an AC signal amplitude of 998.5 mV. The bias voltage was swept from accumulation to inversion operation mode and vice-versa in order to determine the hysteresis of the C-V curve. The hysteresis in the C-V curves is a measure of another non-ideal condition in MOS capacitors, associated to origin of the mobile charge density ($N_\text{m}$), due to the trapping of charges induced in the oxide layer in the fabrication steps and contamination during the irradiation campaigns~\cite{Fowkes, Repace}. 
As shown in Figure~\ref{fig:2}b, $V_\text{fb}$ for samples scanned from accumulation to inversion conditions, is extracted from C-V curves by differentiating the ($C$/$C_\text{ox}$)$^{-2}$ versus bias voltage curve twice. The gate voltage value corresponding to the peak in the double differentiated curve is equal to the $V_\text{fb}$ value. Usually, second differentiation introduces a great deal of noise as the bias voltage sweeps toward the deep-depletion region but that can be avoided by smoothing the data~\cite{Schroder}. $N_\text{f}$ is determined using the following relation: 
\begin{equation}
\label{eq:1}
N_\text{f} = \frac{\Delta V_\text{fb} \times C_\text{ox}}{eA}
\end{equation}
 where, $e$ is the fundamental unit of charge, $A$ refers to the cross-sectional area of the gate and $C_\text{ox}$ is the capacitance value estimated from the accumulation region of the C-V plots. In a similar manner effective concentration of mobile charge density ($N_\text{m}$) can be determined using the following relation: 
\begin{equation}
N_\text{m} = -  \frac{\Delta V_\text{hysteresis} \times C_\text{ox} }{eA}
\label{mobile}
\end{equation}

where $\Delta V_\text{hysteresis}$ is the hysteresis in the flat-band voltage attained while sweeping the gate voltage from accumulation to inversion region and vice-versa. 
It is evident in Figure~\ref{fig:2}c that the effective concentration of fixed oxide charges increases by a factor of $\sim$2.44 on irradiation up to doses of 1~MGy for samples with an additional dielectric layer of HfO$_2$ along with Al$_2$O$_3$. As the concentration of the negative charge accumulated in the oxide-silicon substrate interface increases with gamma irradiation, it leads to energy band bending further down in equilibrium. Therefore, in order to attain flat-band condition, the gate bias has to be swept towards higher positive bias voltage values in order to overcome the inherent bending of the bands. Figure~\ref{fig:2}d shows that $N_\text{m}$ values increase with gamma dose. MOS samples with Al$_2$O$_3$ + HfO$_2$ show higher sensitivity towards gamma irradiation up to a dose of 676~kGy as the effective $N_\text{m}$ values increases by a factor of $\sim$2.3 more than that for samples with Al$_2$O$_3$ alone as dielectric layer. Further, on irradiating the MOS samples up to 1~MGy, devices with Al$_2$O$_3$ alone as field insulator layer were susceptible to an early breakdown. 
 % A negative charge in the oxide near the p-type silicon substrate aids in increasing the holes on the surface, so the gate voltage required to invert the surface to become n-type is increased, as a consequence of which, this behavior will lead to a shift in the threshold voltage as well to higher positive values. 
\section{Radiation Hardness Studies of Gamma Irradiated Detectors}
\label{sec:radiationhardness}

\subsection{I-V/C-V measurements of Gamma-irradiated PIN Diode Structures}
\label{sec:IV_CV_measurements}
\begin{figure}[h!]
\centering % \begin{center}/\end{center} takes some additional vertical space
\includegraphics[width=1\textwidth]{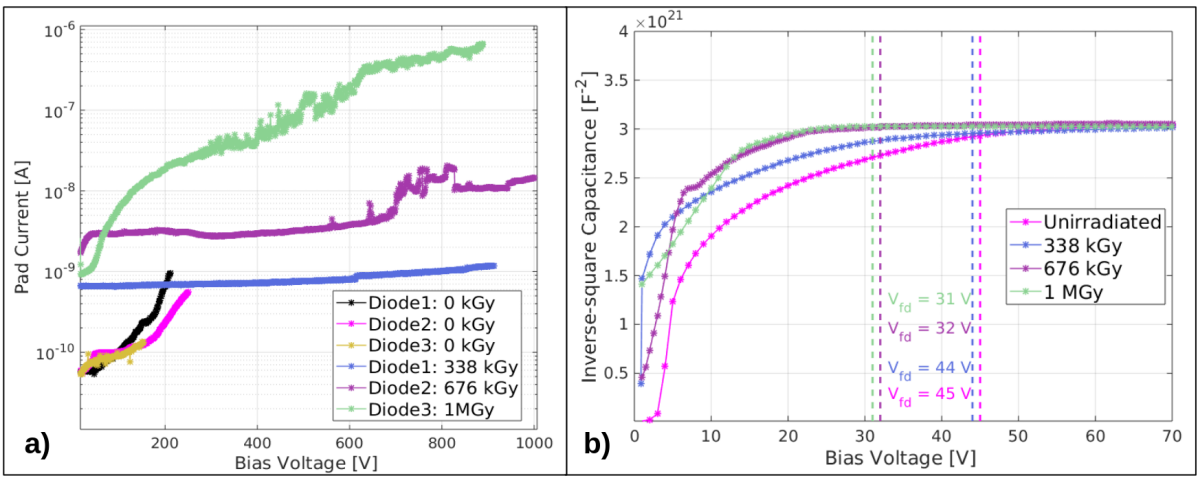}
% "\includegraphics" from the "graphicx" permits to crop (trim+clip)
% and rotate (angle) and image (and much more)
\caption{\label{fig:4} \textbf{a)~}I-V and \textbf{b)~}C-V measurements of non-irradiated and gamma-irradiated PIN diodes with Al$_2$O$_3$ as field insulator. C-V measurements were performed at a frequency of 10~kHz.}
\end{figure}

Figure~\ref{fig:4}a shows current-versus-voltage (I-V) curves of non-irradiated and gamma-irradiated PIN diode prototype with Al$_2$O$_3$ as the dielectric material that were measured at -20$^{\circ}$C. The leakage current shown in the figure corresponds to the values read out from the active pad region consisting of the pn-junction. The first guard ring surrounding the pad was grounded. The non-irradiated samples possess low leakage currents of a magnitude of a few $\sim$100~pA. However, the edge effects of the sensors contribute to a high total current, thereby terminating the measurements as the current value reached its compliance of 50~\textmu A. Interestingly, the compliance is not attained for gamma-irradiated sensors. The gamma-irradiated sensors could be biased up to a voltage of $\sim$900~V before the total current reaches a compliance value of 1~mA. The pad current values for sensors irradiated up to doses of 338~kGy and 676~kGy is increased by a factor of approximately 1.25 and 3, respectively; while the pad current value for the sensor irradiated at 1~MGy is increased by $\sim$2 orders of magnitude. Figure~\ref{fig:4}b shows inverse-squared capacitance versus bias voltage plots derived from C-V measurements performed at a frequency of 10~kHz on the non-irradiated and gamma-irradiated PIN diodes. The full depletion voltage ($V_\text{fd}$) of the non-irradiated sensors is $\sim$50~V. The $V_\text{fd}$ value does not vary on irradiating the sensor up to 338~kGy. However, $V_\text{fd}$ value is reduced by $\sim$20~V on irradiating up to doses of 676~kGy and 1~MGy.  This corresponds to an effective doping concentration value of 6.69~$\times$~10$^{11}$~cm$^{-3}$ for non-irradiated sensor that subsequently gets reduced by a factor of 1.71 on irradiating it at 1~MGy. 

\subsection{Transient Current Technique (TCT) measurements}
\label{sec:TCT}

\begin{figure}[tbph!]
\centering % \begin{center}/\end{center} takes some additional vertical space
\includegraphics[width=1\textwidth]{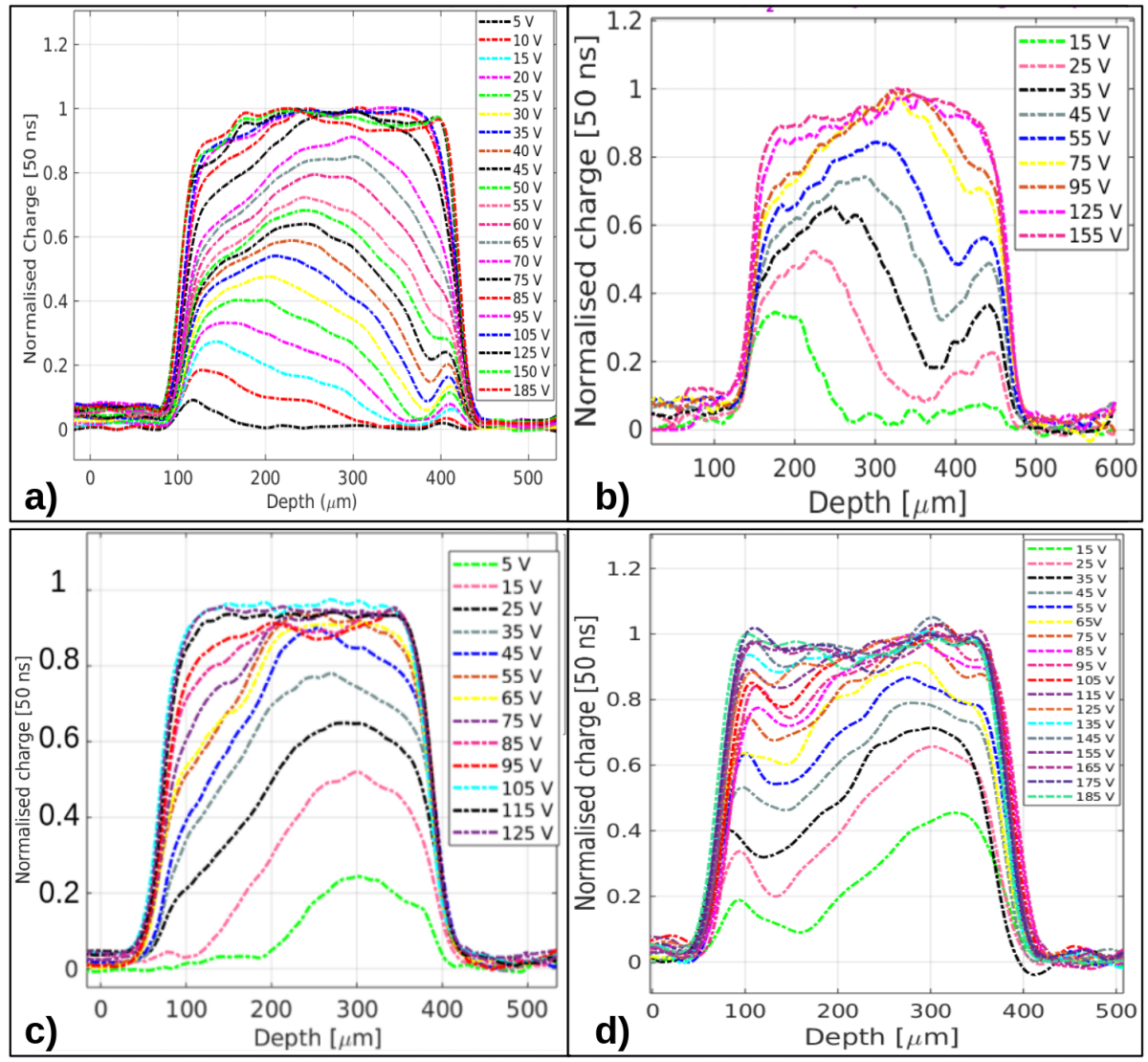}
% "\includegraphics" from the "graphicx" permits to crop (trim+clip)
% and rotate (angle) and image (and much more)
\caption{Charge collection versus depth profiles at varying bias voltages of \textbf{a)}~non-irradiated and gamma-irradiated sensors at doses of \textbf{b)}~338~kGy, \textbf{c)}~676~kGy and \textbf{c)}~1~MGy measured using e-TCT. The left and right-hand sides of the profiles refer to the front and rear sides of the sensor, respectively. All the irradiated sensors were measured at a temperature of -20$^{\circ}$C.}
\label{SPA}
\end{figure}

\begin{figure}[tbph!]
\centering % \begin{center}/\end{center} takes some additional vertical space
\includegraphics[width=1\textwidth]{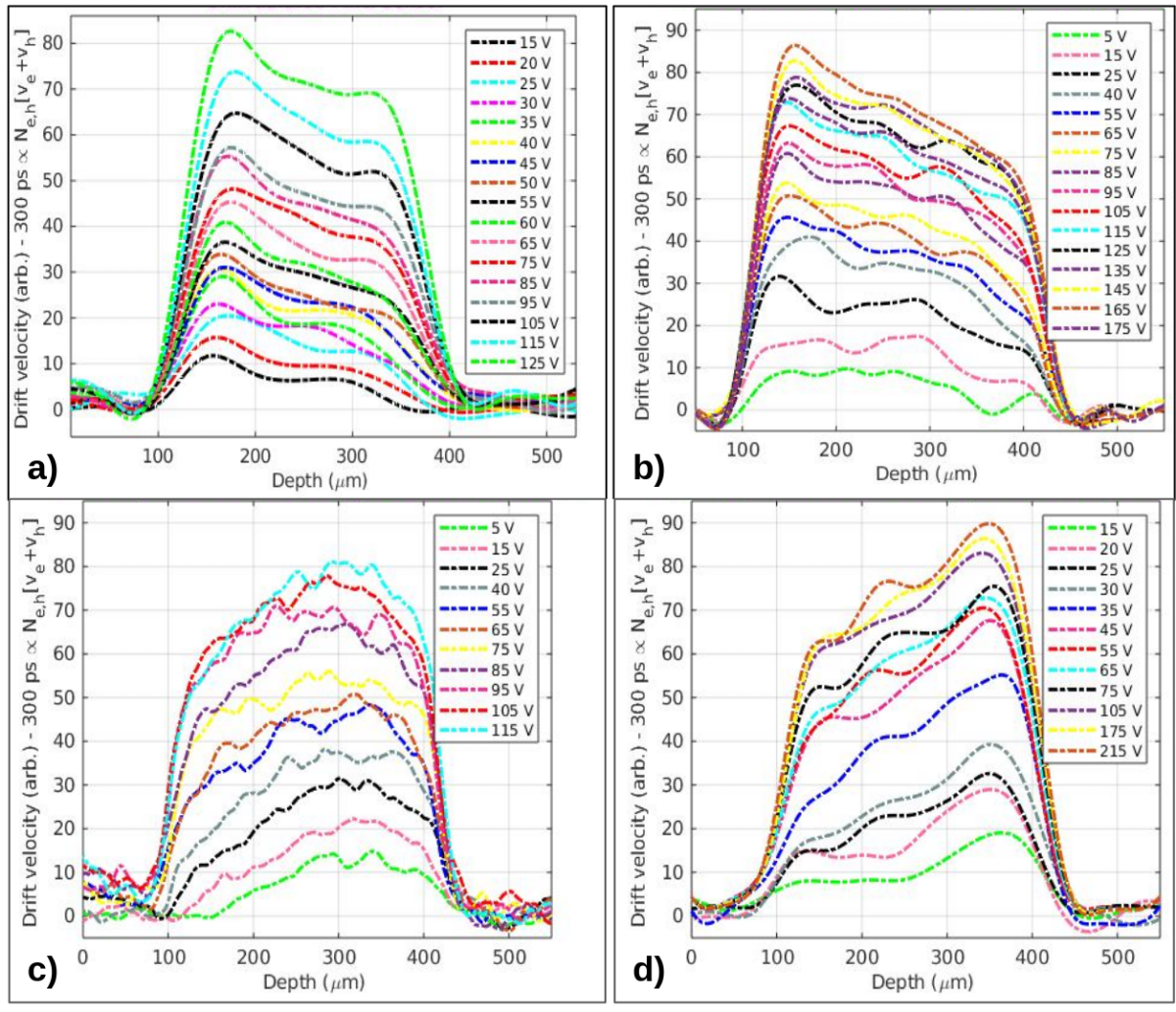}
% "\includegraphics" from the "graphicx" permits to crop (trim+clip)
% and rotate (angle) and image (and much more)
\caption{Drift velocity versus depth profiles at varying bias voltages for sensors \textbf{a)}~non-irradiated and irradiated with gamma-rays up to dose values of \textbf{b)}~338~kGy, \textbf{c)}~676~kGy and \textbf{d)}~1~MGy measured using e-TCT. The left and right-hand sides of the profiles refer to the front and rear sides of the sensor, respectively. All the irradiated sensors were measured at a temperature of -20$^{\circ}$C.}
\label{electricfield_SPA_TCT} 
\end{figure}

Single Photon Absorption (SPA)-TCT is an electrical characterisation method that makes use of a fast pulse infrared (IR) laser imitating a minimum ionising particle (MIP). The photons of the IR laser traverse through the active region of the detector thereby leading to a continuous energy deposition along the beam direction that provides a close approximation to high-energy particles interacting with the detector and the signal read-out is due to the charge carriers that are generated within the bulk of the depleted region. According to the Shockley-Ramo theorem~\cite{Ramo}, the resulting transient signal detected by the oscilloscope provides information about the movement of the charge carriers due to the electric field within the sensor. The collected charge across the depleted bulk can be investigated by integrating the induced current signal over time as the IR laser beam is scanned across the depth of the detector. This methodology of projecting the laser beam across the edge of the sensor to study the charge collection and drift velocity profiles is called edge-TCT (e-TCT). Moreover, the electric field inside a PIN diode can be studied from the drift velocity, which itself can be extracted from the waveform using the prompt current method~\cite{Canali}. 

%Similar charge collection and drift velocity profile measurements were performed across the depth of the active region of the sensors at varying bias voltage values using Two-Photon Absorption - TCT (TPA-TCT). The TPA-TCT uses infrared laser with wavelength inside the quadratic absorption regime of silicon, to only generate excess charge carrier by TPA. Compared to SPT-TCT, the light absorption is not continuous along the beam propagation direction, but excess charge carriers are only generated inside a confined volume around the laser's focal point, allowing to probe the bulk of the detector with a three dimensional resolution.
%he drift velocity profiles can be predicted by integrating over $\sim$~300~ps of initial rise time of the current signal as the maximum drift distance of the charge carriers within this time frame is approximately 30~\textmu m, comparable to the beam size. This is commonly known as prompt current method

%\begin{figure}[h!]
%\centering % \begin{center}/\end{center} takes some additional vertical space
%\includegraphics[width=1\textwidth]{TPA_TCT_plots_nonirrad.png}
% "\includegraphics" from the "graphicx" permits to crop (trim+clip)
% and rotate (angle) and image (and much more)
%\caption{\label{TPA_TCT} \textbf{a)}~Charge collection versus depth and \textbf{b)}~drift velocity versus depth profiles of a non-irradiated sensor at varying bias voltages measured using TPA-TCT.}
%\end{figure}

The full depletion voltage for the non-irradiated device, based on the charge collection profiles, was estimated to be approximately close to 95~V using e-TCT with SPA, as shown in Figure~\ref{SPA}a. The attained $V_\text{fd}$ value analysed in e-TCT measurements was $\sim$45~V higher than the value extracted from the C-V measurements for non-irradiated and irradiated sensors. The reason behind this feature was the high voltage filter connected to the bias line that supplies a lower bias value to the device under test compared to the actual applied voltage from the Keithley, due to the high total leakage current read out from the sensors. %However, as shown in Figure~\ref{TPA_TCT}a, the full depletion voltage value for the non-irradiated sensor was observed at $\sim$~40~V, when the guard rings were left floating. This value is coherent to $V_\text{fd}$ extracted from C-V measurements in Figure~\ref{fig:4}b. Additionally, a second peak in the charge collection profiles close to the ohmic contact was observed as an artifact in the measurements due to reflection of the beam from the metallic structures at the back-plane into the active volume of the detector. These focused reflections lead to unwanted absorption of light in TPA-TCT~\cite{Wiehe, Pape}.
% Nevertheless, diffused scattering of the photons and divergent reflections do not lead to the creation of additional charge carriers

Further, the charge collection profiles as shown in Figures~\ref{SPA}b, \ref{SPA}c and \ref{SPA}d show that the full depletion voltage value is reduced by $\sim$20~V,when irradiating the samples up to 1~MGy. The charge collection profiles are coherent to the drift velocity profiles of irradiated sensors, shown in Figure~\ref{electricfield_SPA_TCT}c and \ref{electricfield_SPA_TCT}d, that demonstrate space charge sign inversion of the bulk as the depletion initiates from the back-plane. This feature gets more prominent on samples irradiated at a dose of 1~MGy in comparison to the ones irradiated at 676~kGy. No significant change in the drift velocity profiles were observed in a non-irradiated sensor and a detector irradiated up to an accumulated dose of 338~kGy, as analysed in Figure~\ref{electricfield_SPA_TCT}a and \ref{electricfield_SPA_TCT}b. This suggests that the depletion in unirradiated devices and in samples irradiated up to a dose value of 338~kGy occurs starting from the n$^+$-implants located at the top of the sensors.

\subsection{Inter-pixel resistance measurements for non-irradiated and gamma-irradiated AC coupled pixel detectors.}
\label{interpixel_resistance}
\begin{figure}[h!]
\centering % \begin{center}/\end{center} takes some additional vertical space
\includegraphics[width=0.5\textwidth]{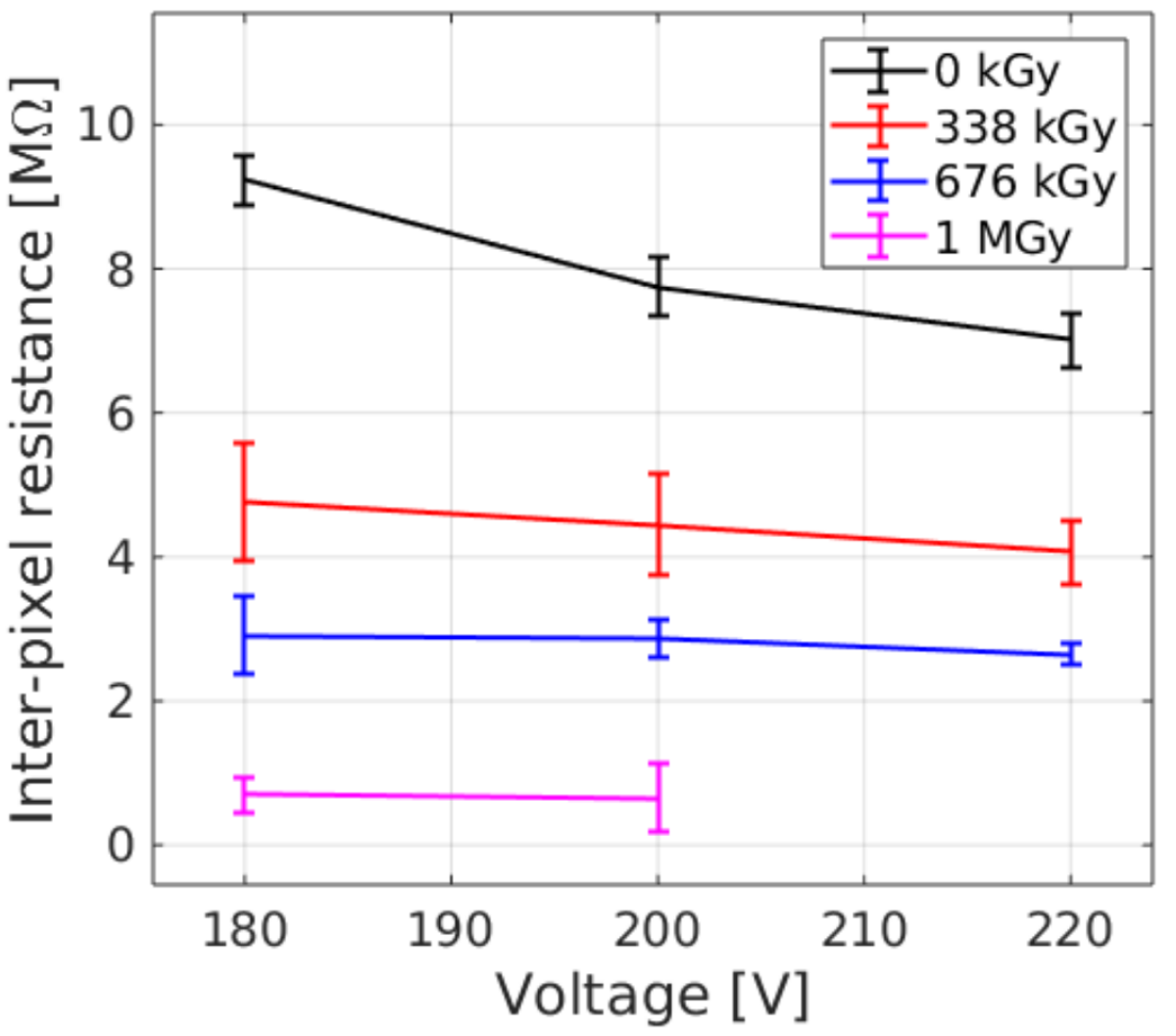}
% "\includegraphics" from the "graphicx" permits to crop (trim+clip)
% and rotate (angle) and image (and much more)
%\caption{\label{interpixel_resistance}\textbf{a)}~Leakage current versus bias voltage curves for inter-pixel resistance measurements determined by recording the current values for an individual pixel in the matrix by grounding the surrounding adjacent pixels in two different configurations in case of non-irradiated and irradiated AC-coupled pixel sensors with Al$_2$O$_3$ as the field insulator.  \textbf{b)}~Variation of inter-pixel resistance values measured for AC-coupled pixel detectors irradiated with Co-60 gamma radiation doses of 1~MGy.}
\caption{\label{interpixel_resistance}~Variation of inter-pixel resistance values measured for AC-coupled pixel detectors irradiated with Co-60 gamma ray source up to dose of 1~MGy.}
\end{figure}

Inter-pixel resistance was measured at bias voltage values from 180--220~V for non-irradiated and gamma-irradiated AC-coupled pixel sensors. During the measurements, the current was read out from the DC-pad of an arbitrary pixel in the centre of a 3~$\times$~3 matrix while the adjacent pixels surrounding it were grounded. The inter-pixel resistance value of non-irradiated devices was of a magnitude $\sim$10~M$\Omega$, equivalent to DC-coupled pixel sensors with same pitch size~\cite{Bonvicini, Alam}. As shown in Figure~\ref{interpixel_resistance}, the magnitude of inter-pixel resistance value was reduced by a factor of 6 on irradiating the sensors to a dose of 1~MGy as the leakage current increases by $\sim$2 orders of magnitude due to radiation induced damage. The lower limit of the inter-pixel resistance is given by the requirement of preventing a significant signal distribution to the neighboring channels within a typical shaping time of 25~ns. This leads to a lower limit of the resistance of about 1~M$\Omega$ in case of p-stop and p-spray implemented as pixel isolation strategies. The measured inter-pixel resistance values for detectors with Al$_2$O$_3$ as field insulator, irradiated up to a dose of 1~MGy, lie within this minimum requirement.

\section{Conclusion and Discussions}

Characterization of MOS devices with high-$K$ negative oxide indicates that a negative charge accumulation is induced by gamma irradiation based on the study of the flat-band voltage. The negative oxide charge during the irradiation is an essential prerequisite of radiation hardness resiliency of n$^+$/p$^-$/p$^+$ (n-on-p) particle detectors widely intended to be used in future high-luminosity experiments. MOS devices with Al$_2$O$_3$ + HfO$_2$ possess higher sensitivity to gamma irradiation in comparison to samples with Al$_2$O$_3$ alone as an insulating layer.

A reduction in full-depletion voltage value and space charge sign inversion was observed in sensors irradiated with gamma rays up to a dose of 1~MGy. Additional investigation is ongoing using Deep-Level Transient Spectroscopy (DLTS) and Thermally Stimulated Current (TSC) methods to understand the defects introduced in the sensors that could possibly explain the reason for space charge sign inversion observed in e-TCT measurements. However, the inter-pixel resistance values show promising results for both non-irradiated and gamma irradiated AC-coupled pixel sensors with high-$K$ negative oxide as field insulator as the pixels are electrically isolated from each other, even for heavily irradiated devices. 
%\paragraph{Up to paragraphs.} We find that having more levels %usually
%reduces the clarity of the article. Also, we strongly discourage the
%use of non-numbered sections (e.g.~\texttt{\textbackslash
%  subsubsection*}).  Please also see the use of
%``\texttt{\textbackslash texorpdfstring\{\}\{\}}'' to avoid warnings
%from the hyperref package when you have math in the section titles
%Further, DLTS measurements of p-type MCz-Si with high resistivity show defects arising due to thermal double donors, high concentration of oxygen nuclei that precipitates during the silicon crystal growth thereby contributing its susceptibility to Fe contamination and creation of C-O-H complexes due to incorporation of hydrogen during chemical etching with BHF.

\acknowledgments

S. Bharthuar would like to acknowledge the Magnus Ehrnrooth Foundation for financial support. Facilities for detector fabrication were provided by Micronova Nanofabrication Centre in Espoo, Finland. The samples were irradiated with Co-60 gamma ray source at the Laboratory for Radiation Chemistry and Dosimetry (LRKD), in Ruđer Bošković Institute, Croatia. The C-V, I-V, and e-TCT measurements were performed at the Detector Laboratory in Helsinki Institute of Physics. %The authors are grateful for Dr.~Michael Moll’s assistance and guidance in proceeding with the DLTS and TPA-TCT measurements in EP-DT-DD SSD lab at CERN.

%\paragraph{Note added.} This is also a good position for notes added
%after the paper has been written.


\end{document}